\documentclass[sigconf]{acmart}

\usepackage[ruled,linesnumbered,noend]{algorithm2e}
\usepackage{multirow}
\usepackage{xcolor}
\usepackage{subfigure}

\allowdisplaybreaks[1]

\AtBeginDocument{%
  \providecommand\BibTeX{{%
    \normalfont B\kern-0.5em{\scshape i\kern-0.25em b}\kern-0.8em\TeX}}}

\copyrightyear{2021} 
\acmYear{2021} 
\setcopyright{acmcopyright}\acmConference[ICPP '21]{50th International Conference on Parallel Processing}{August 9--12, 2021}{Lemont, IL, USA}
\acmBooktitle{50th International Conference on Parallel Processing (ICPP '21), August 9--12, 2021, Lemont, IL, USA}
\acmPrice{15.00}
\acmDOI{10.1145/3472456.3473513}
\acmISBN{978-1-4503-9068-2/21/08}

\begin{document}

\title{Dubhe: Towards \underline{D}ata \underline{U}n\underline{b}iasedness with \underline{H}omomorphic \underline{E}ncryption in Federated Learning Client Selection}


\author{Shulai Zhang}
\affiliation{%
  \institution{Shanghai Jiao Tong University}
  \country{China}
  }
\email{zslzsl1998@sjtu.edu.cn}

\author{Zirui Li}
\affiliation{%
  \institution{Shanghai Jiao Tong University}
  \country{China}
  }
\email{suffix_array@sjtu.edu.cn}

\author{Quan Chen}
\affiliation{%
  \institution{Shanghai Jiao Tong University}
  \country{China}
  }
\email{chen-quan@cs.sjtu.edu.cn}

\author{Wenli Zheng}
\affiliation{%
  \institution{Shanghai Jiao Tong University}
  \country{China}
  }
\email{zheng-wl@cs.sjtu.edu.cn}

\author{Jingwen Leng}
\affiliation{%
  \institution{Shanghai Jiao Tong University}
  \country{China}
  }
\email{leng-jw@cs.sjtu.edu.cn}

\author{Minyi Guo}
\affiliation{%
  \institution{Shanghai Jiao Tong University}
  \country{China}
  }
\email{guo-my@cs.sjtu.edu.cn}

\renewcommand{\shortauthors}{Shulai Zhang, et al.}

\begin{abstract}
Federated learning (FL) is a distributed machine learning paradigm that allows clients to collaboratively train a model over their own local data. FL promises the privacy of clients and its security can be strengthened by cryptographic methods such as additively homomorphic encryption (HE). However, the efficiency of FL could seriously suffer from the statistical heterogeneity in both the data distribution discrepancy among clients and the global distribution skewness. We mathematically demonstrate the cause of performance degradation in FL and examine the performance of FL over various datasets. To tackle the statistical heterogeneity problem, we propose a pluggable system-level client selection method named {\bf Dubhe}, which allows clients to proactively participate in training, meanwhile preserving their privacy with the assistance of HE. Experimental results show that Dubhe is comparable with the optimal greedy method on the classification accuracy, with negligible encryption and communication overhead.
\end{abstract}


\keywords{federated learning, client selection, homomorphic encryption}


\maketitle

% \clearpage
\section{Introduction}

\begin{figure}
\centering
\includegraphics[width=0.4\textwidth]{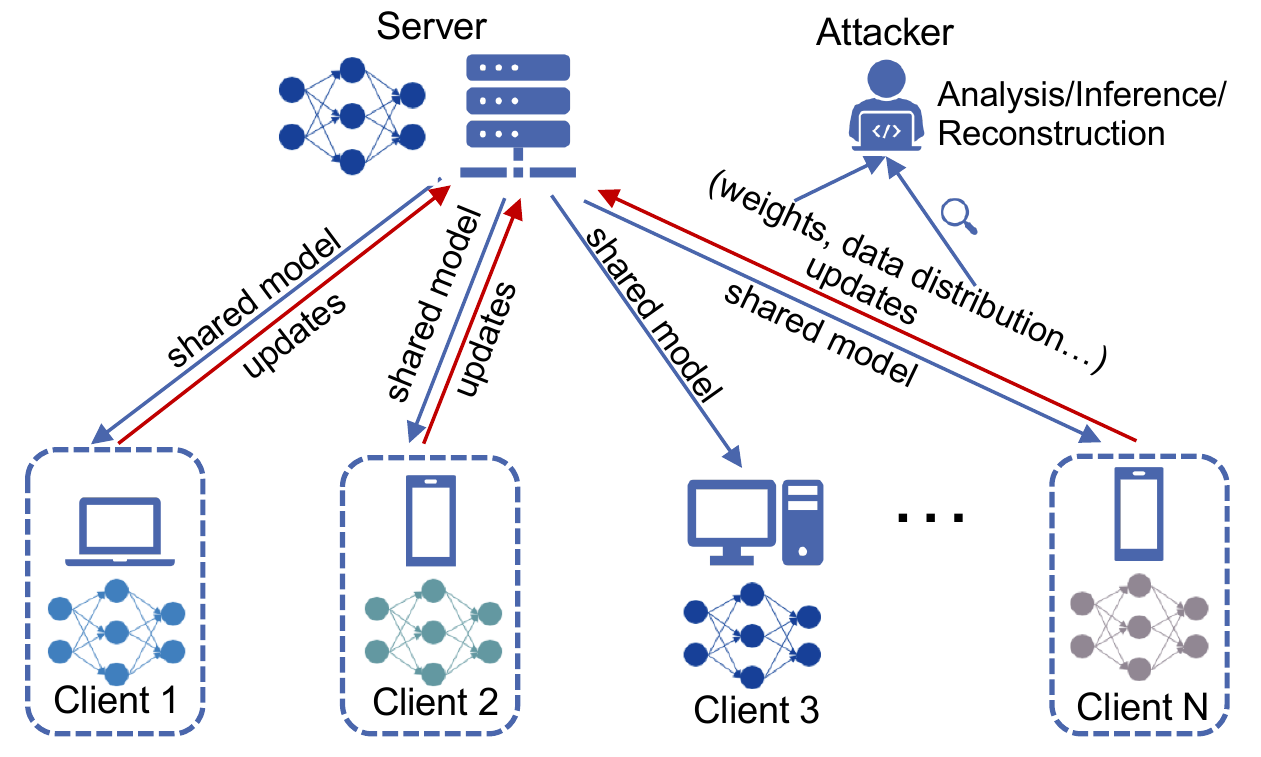}
\caption{The framework of Federated Learning. Clients in dashed boxes are selected and participate in training in one specific round. Attackers are able to infer a client's information from its updates.}
\label{fig:picture000}
\vspace{-4mm}
\end{figure}

In typical distributed machine learning systems, datasets are separated and distributed into homogeneous computing devices in an unbiased and independent and identically distributed (IID) manner to speed up computation. Nevertheless, in the other bottom-up structure, data are generated and collected in a non-identically distributed (non-IID) manner across the network by heterogeneous devices or parties, bringing great concern of data privacy~\cite{li2019survey, truex2019hybrid}. With decentralized computing devices and parties which are referred to as clients gaining usable computing resources, federated learning (FL), first proposed by Google~\cite{mcmahan2017communication}, is becoming a feasible solution to train a global model without breaking the privacy demand. In this paradigm as shown in Figure~\ref{fig:picture000}, a portion of clients are selected in each training round. Each client locally performs a few iterations of training and only sends model updates to the central server. Then the global model is obtained by aggregating the collected local model updates, without private data leakage.
%By encrypting model parameters and then executing the add operation over the encrypted space, Homomorphic Encryption (HE) can eradicate the reverse inference attempts with the price of largely increased data transferred.

In situations where clients are massive mobile or IoT devices (e.g., up to $10^{10}$) \cite{chai2020tifl} in FL, it is unrealistic to achieve full client participation. 
In practice, only a small fraction of clients participate in each training round, bringing the problem of client selection. 
Autonomy and privacy are two primary principles in designing a client selection algorithm. 
Clients in FL are often autonomous and under independent control~\cite{li2019survey}, so each client is supposed to decide whether to participate in each training round. 
The privacy demand requires the selection method to maintain the privacy restrictions and avoid improving performance with the price of privacy.

Client selection methods can alleviate the performance degradation caused by the heterogeneity in FL. 
The storage, computational, and communication capabilities of each device in a FL system may differ due to variability in hardware, network connectivity and power, forming the system heterogeneity. 
Client selection strategies considering system heterogeneity can avoid stragglers and take full advantage of restricted system resources. Statistical heterogeneity is another source of heterogeneity, originated from the non-IID data collected by clients. In a FL system that has a monotonic training goal, there is a tug-of-war among different clients, with each client pulling the model to reflect its own data \cite{hsieh2020non}. The adverse effects of statistical heterogeneity can be exacerbated when the partial client participation is biased from the full client participation, sharply accelerating the error convergence \cite{cho2020client}. Moreover, the skewness in global data distribution is also fundamental and pervasive in FL, making the client participation much more biased, especially when clients are randomly selected in each round. Thus, there is an urgent need for client selection strategies that promise data unbiasedness in FL.
%Definitely, an optimal data balancing solution can be reached knowing all clients' data distribution [], while inevitably breaking the privacy-preserving rule unfortunately.

%\added[]{A diagram to show the general view of FL? and show where the information an be leaked, and what information can be leaked.}

%\added[]{client selection, huge amount of device, small participation, client selection, autonomy, privacy, \\ system heterogeneity, statistical heterogeneity, non-IID, tug-of-war, biased participation, global skewness (a problem in centralized ML)}

%\added[]{privacy, data information leakage, homomorphic encryption}
Client information can be leaked from any information transmitted between clients and the server as shown in Figure~\ref{fig:picture000}, including weights and data distribution. Without additional mechanisms, malicious clients/servers still have possibilities to reconstruct the local data or operate the ``membership attack''~\cite{niknam2020federated} from the public shared models in a FL system. Once the data distribution of any client is leaked, GAN(Generative Adversarial Network)-based methods~\cite{wang2019beyond, hitaj2017deep} are able to reconstruct original data based on the data distribution, thereby weakening the differential privacy requirement~\cite{geyer2017differentially, truex2019hybrid}. As a simple explanation, the leakage of label distributions of users' data can directly give away personal flavors. Thus, there is still a great demand for algorithms to ensure no information leakage throughout FL, among which additively homomorphic encryption (HE)~\cite{paillier1999public}, notably the Paillier cryptosystem, has been widely used in current secure FL frameworks such as FATE~\cite{FATE}. With additively HE, any additive operations can be performed directly on ciphertexts, without decrypting them in advance. Thus, no information can be learned by any external party (including the server) during data transmission and aggregation.

Based on additively HE, we propose {\bf Dubhe}, a proactive client selection system to achieve data unbiasedness without data information leakage. 
In Dubhe, each client enjoys the maximum autonomy, since each client decides whether to participate in each training round by its own calculated probability. 
The probability calculation method is shared by all clients, as a function of each client's own data distribution and the global data distribution. 
In each round, the data distribution is transmitted between clients and the server through a well-designed encrypted vector, named registry in Dubhe, to guide the decisions of clients. The data unbiasedness, as well as the model accuracy is able to be further improved by tentative multi-time client selections before training. The main contributions of this work are as follows:

\begin{itemize}
    \item %We mathematically derives the weight divergence in FL. 
    We mathematically prove that the weight divergence can be evaluated from the data distribution discrepancy among clients, and the biased client participation caused by global data imbalance. %We generate datasets with statistical heterogeneity quantitatively according to the two dimensions.
    \item We design a pluggable client selection method based on additively HE. The method achieves data unbiasedness, ameliorating the effect of global data imbalance and increasing the model test accuracy. %We find that the effect of Dubhe can be enhanced by repeating the selection process tentatively.
    \item %\textcolor{black}{Analyzing the impact of system parameters or overhead is not a contribution. The contribution could be the new method, new idea, new design etc. Please find one more novel method in Dubhe. Probably the multi-time selection process???}
    We propose a multi-time selection method for client determination to further balance data in each round. The multi-time selection is also utilized in the parameter search process to enhance the precision of Dubhe.
    %We further analyze the impact of system parameters on the effect of data balancing and point out the applicable conditions of parameters. We also analyze the encryption and communication overhead of Dubhe.
\end{itemize}

Dubhe improves the accuracy by $58.7\%$ on {\color{black} MNIST~\cite{lecun1998mnist}, $48.1\%$ on CIFAR10~\cite{krizhevsky2009learning}, and $84.4\%$ on FEMNIST~\cite{caldas2018leaf}} compared with the optimal greedy method. With the multi-time selection process, the accuracy can be improved by another $69.5\%$ (MNIST) and $18.8\%$ (CIFAR10). 

%Compared with the transferred model updates, the encrypted information in Dubhe has a size reduction by orders of magnitude.

\section{Background and Related Work}
In this section, we introduce the related work in the functionality (client selection), the dilemma (statistical heterogeneity), and the property (security promise) in Dubhe.

\subsection{Client selection in FL}
Smart client selection methods are able to solve existing problems in FL. FedCS in \cite{nishio2019client} actively manages the resources of heterogeneous clients by grouping clients based on their hardware and wireless resources in order to save communication resources.
There are client selection methods aiming to improve the training performance in FL. FAVOR~\cite{wang2020optimizing} uses a deep Q-learning model to select clients to maximize a reward that encourages the increase of accuracy and penalizes the use of more rounds. The framework in Astraea~\cite{duan2020self} acknowledges the access of data distribution of clients. It uses a greedy algorithm to balance data to reach the optimum in a global data imbalance setting. Y. J. Cho et.al~\cite{cho2020client} and J. Goetz et.al~\cite{goetz2019active} propose to select clients based on the local loss of the global model on each client. 

However, computations of the local loss are performed on clients each round, which is an additional computation burden of clients. {\bf In all above proposed schemes, the autonomy of clients is not taken into consideration.}

%\added[]{do not cite the paper in \cite{goetz2019active} and \cite{cho2020client} this way. You can use either the system name, or the authors name as the citation. Like: Astraea~\cite{duan2020self}, Hsieh et. al~\cite{reddi2020adaptive}, xxxx[xx]}

\subsection{Statistical heterogeneity in FL}

The statistical heterogeneity is explicit in local data skewness, data discrepancy among clients, and global data skewness. 
%There are numerous algorithms designed to solve the statistical heterogeneity in FL. Most of the algorithms are focusing on the loss design[], the optimizer design [], and the aggregation methods [], mostly at an algorithm level. There are also specific research on the analysis of these algorithms on a systematic level [].

{\bf Local data skewness:}
The skewness in local data distribution is a natural problem and there are numerous applicable techniques including sampling \cite{chawla2002smote} and cost-sensitive learning methods \cite{thai2010cost} to ameliorate data skewness. However, there are constraints to apply these tools in FL because there can exist data absence in some classes in local datasets. Y. Zhao et.al~\cite{zhao2018federated} proposed to share a public dataset to all clients to alleviate the problem. A new loss function called Ratio Loss is introduced in \cite{wang2020towards} to mitigate the effect of data imbalance in FL.

{\bf Data discrepancy among clients:} There are some algorithm-level solutions to ameliorate Non-IID data distribution among clients. FedProx \cite{li2018federated} constrained the divergence between the local model and the global model by introducing a L2 regularization term in the local objective function. FedNova~\cite{wang2020tackling} gave larger weights to clients who conduct more local epochs in the aggregation stage. SCAFFOLD~\cite{karimireddy2020scaffold} introduced control variates for the server and clients and corrects the local updates by adding the drift in the local training. 
%S. Reddi et. al~\cite{reddi2020adaptive} considered the aggregation also a SGD process to reduce the effect of client heterogeneity.

{\bf Global data skewness:}
The global data distribution skewness problem is pervasive in FL~%\cite{hsieh2020non, luo2019real, hsu2020federated}
\cite{hsieh2020non, hsu2020federated}. K. Hsieh et. al~\cite{hsieh2020non} emphasized this problem through the analysis on a real-world dataset. %and they imply that group normalization can recover much of the accuracy loss of batch normalization in this situation. 
Datasets used in \cite{hsu2020federated} also revealed significant global skewness and two algorithms (FedIR and FedVC) that intelligently re-sample and re-weight over the client pool are used to stabilize the training. 

%\added[]{However, what is the weakness??}
%The datasets are often quantified by various forms of distances (KL divergence[], Earth Move Distance (EMD)[]) among clients.
To the best of our knowledge, no contribution has been devoted to compensating the statistical heterogeneity within the client selection stage in the existing literature of secure FL.

\subsection{Security in FL}
In this work, we assume that the server is honest-but-curious, which is a commonly used threat model in the existing FL literature. The server always attempts to recover clients' information based on legally obtained data.

Without additional mechanisms, a malicious server has many possibilities to reconstruct local data. Differential privacy, secure multi-party computation, and homomorphic encryption (HE) are widely used in FL to protect client privacy, among which HE allows computation to be performed directly on ciphertexts, thereby protecting clients' information from being obtained by unauthorized parties. HE can be easily plugged into existing FL solutions \cite{liu2020secure, hardy2017private, zhang2020batchcrypt} to augment privacy-preserving. At the moment, Paillier \cite{paillier1999public} is a mature cryptosystem that has already been used in FL systems, notably FATE \cite{FATE}. 

In a secure FL system with HE, a HE key-pair is generated and dispatched to all clients by an agent before iterations. Each participating client uses the public key to encrypt its message and sends the encrypted message to the server. The server aggregates received messages and dispatches the result to all clients. Then each client can decrypt the result with the private key. In the existing literature, the transmitted messages under protection are the updates of models. The advantage of HE enables us to transmit any additional information securely.

However, HE brings high encryption and communication overhead when the plaintext is considerable such as huge model parameters. In practice, the data transfer between clients and the server is extended by approximately $160\times$ compared with directly transferring the plaintext updates and the iteration time is extended by an average of $130\times$ with HE~\cite{zhang2020batchcrypt}.

\section{Motivation of Dubhe}
Random selection aggravates the biased client participation when data among clients are non-IID and the global data distribution is biased. Phenomena from two experiments on a classification dataset CIFAR10 shows that biased client participation caused by random selection degrades the classification accuracy.

\begin{figure}
\centering
\subfigure[Global data skewness in CIFAR10 classification]{
\begin{minipage}[b]{0.45\textwidth}
\includegraphics[width=1\textwidth]{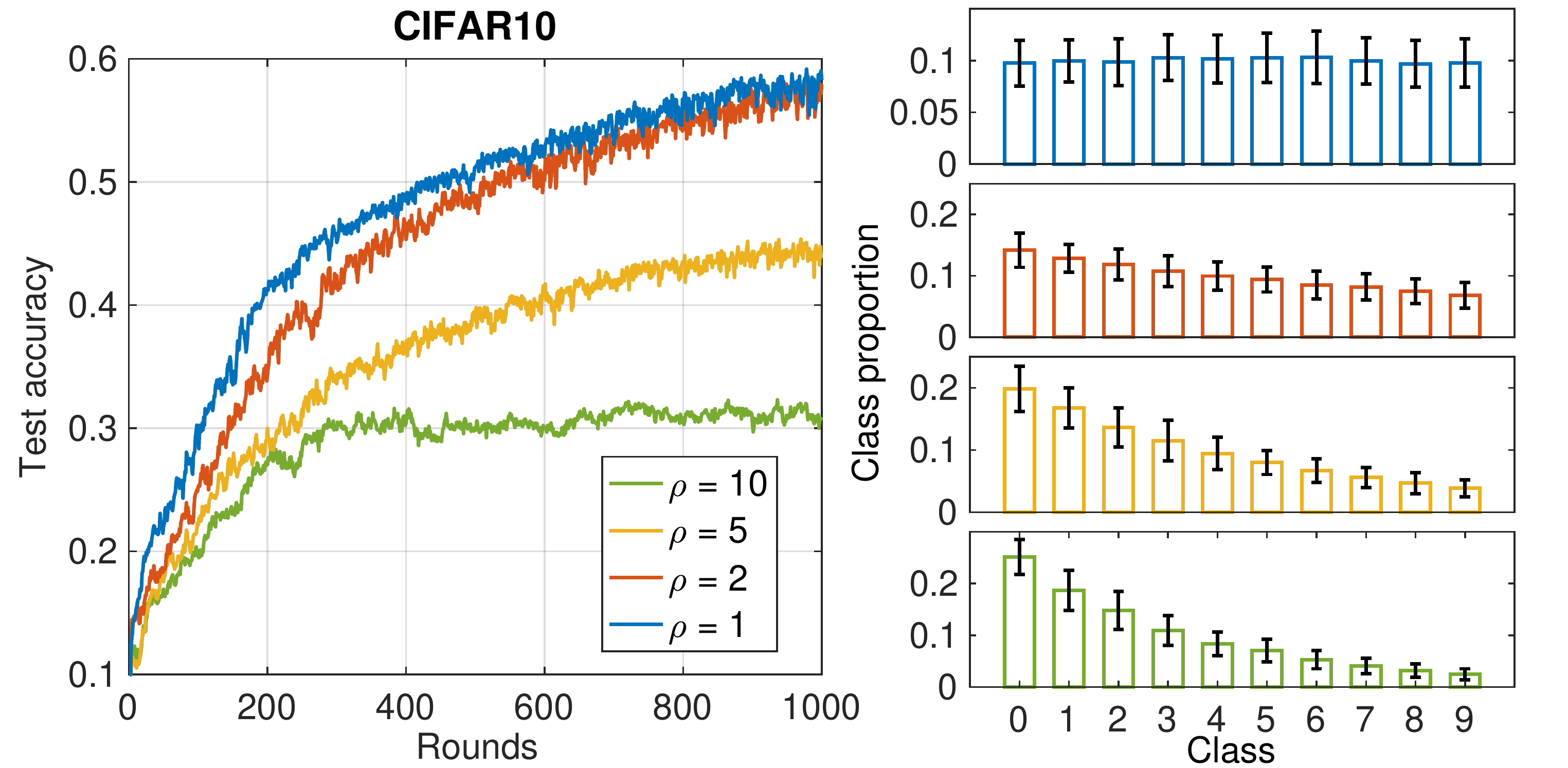}
\vspace{-3mm}
\label{fig:mv_01}
\end{minipage}
}
\vspace{-2mm}
\subfigure[Discrepancy of clients in CIFAR10 classification]{
\begin{minipage}[b]{0.45\textwidth}
\includegraphics[width=1\textwidth]{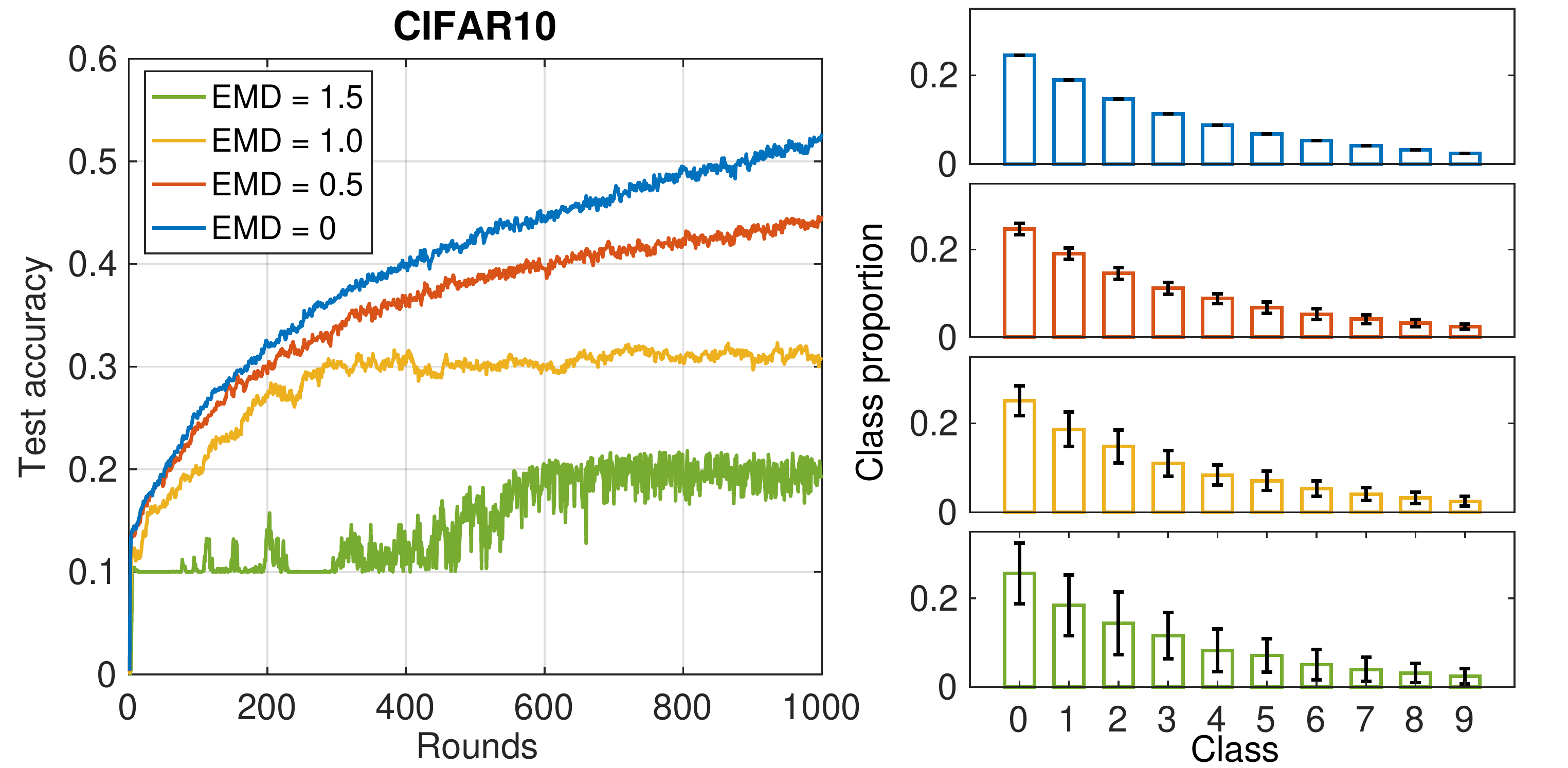}
\vspace{-3mm}
\label{fig:mv_02}
\end{minipage}
}
\vspace{-3mm}
\caption{CIFAR10 classification in different FL settings} 
\label{fig:1}
\vspace{-2mm}
\end{figure}

As shown in Figure~\ref{fig:1}, {\color{black}curves of the top-1 classification accuracy on the balanced test dataset are shown on the left and the expectations of the participated class proportion are shown on the right, colored differently in different cases and with error bars which stand for standard deviations.} The class imbalance ratio $\rho$ is defined as the sample size of the most frequent class divided by that of the least frequent class and is used to evaluate the global data skewness. Larger $\rho$ indicates more skewness. {\color{black} The Earth Mover's Distance (EMD) is the 1-norm distance between two distributions, which are the local data distribution and the global data distribution in our cases specifically. We use the average distance $EMD^{avg}$ to evaluate the data discrepancy among clients. The $EMD^{avg}$ ranges from 0 to 2 and larger values of the $EMD^{avg}$ indicate more differences among clients' data.}
%\added[]{for $\rho$ and EMD, what is the range? Close to 0 is balance or close to 1 is balance?}

In Figure~\ref{fig:mv_01}, the global class proportion is skewed at four different levels and $EMD^{avg} = 1$. It is obvious that with the increase of global data skewness, the test accuracy is decreasing. The results imply that models trained with skewed global data tend to converge at local optimums. As shown in Figure~\ref{fig:mv_02}, the global class proportion is skewed and $\rho\!=\!10$, while the discrepancies among clients are different. As expected, the expectations of the participated class proportion are similar to the global class proportion. 

{\bf The deviation of the participated class proportion from the global class proportion is increasing with the discrepancy among clients, causing the accuracy degradation. We also notice that the variance in participated class proportion will cause fluctuation in training.
}

The negative impact of biased client participation is intuitive and our experiments have shown that the data unbiasedness in each round is able to improve the training performance. Prior researches~\cite{goetz2019active, cho2020client} have comprehended the data balancing issue as the loss minimizing issue, and propose to select clients based on their local loss. 
%However, these researches did not discuss the essence of performance degradation, which is the statistical heterogeneity in this problem. 
However, in these researches, a large proportion of clients are required to be active to conduct additional loss computation in each round, which is equivalent to the training process using all local data without back propagation.
%However, the loss computation is conducted using all local data and this additional computation is actually a full-batch training process without back propagation. It is unreasonable to require clients to 'train' for the loss when they are inactive. 
Astraea \cite{duan2020self} proposed a greedy client selection method to balance data, but with the price of giving away clients' data distribution, which is unacceptable in FL. Besides, the greedy client selection algorithm has a high time complexity which is $O(NK)$, where $N$ is the total client number and $K$ is the participating client number in each round. %When the scale of the FL system increases, the brute-force searching in the greedy client selection becomes the straggler in the FL system in practice.

Therefore, the need for data balancing in each training round and the privacy requirements of clients impel us to develop a smart and light client selection method.

\section{Problem Formulation}
%\added[]{In this section, we XXXX.}
In this section, we model the FL system mathematically. Through mathematical demonstration, we analyze the weight divergence in FL with statistical heterogeneity and illustrate our objective.

\subsection{Modeling a FL system}
In a typical FL system, each client frequently generates and updates the collection of data samples given as a set of input-output pairs. In each round, a subset of all samples in the local dataset is selected and devoted to training. We denote the actual dataset used for training at round $t$ as $D^{(t,k)}$, with size $n^{(t,k)}$.

At each round $t$ the server sends the global model $\omega_t^f$ to all $N$ clients. Then some clients are selected to form a selection pool $\mathcal{S}^t$ at round $t$. Each client $k \in \mathcal{S}^t$ trains the local model $\omega_t^k$ over its local dataset $D^{(t,k)}$ for $m$ epochs. Then the client updates its model with $\omega_{t+1}^k$ and sends it back to the server. The server aggregates the received updates and repeats the above process.

We consider a $C$ class classification problem defined over a compact space $\mathcal{X}$ and a label space $\mathcal{Y}=[C]$, where $[C]=\{1,\cdots, C\}$. The data point $\{ \mathbf{x}^{(i)}, y^{(i)}\}$ from $D^{(t,k)}$ distributes over $\mathcal{X}\times \mathcal{Y}$ following the distribution $p^{(t,k)}$. The distribution $p^{(t,k)}$ follows $p^{(t,k)}(y\!=\!j)=\sum_{\{ \mathbf{x}^{(i)}, y^{(i)}\} \in D^{(t,k)}} \mathbb{I}(y^{(i)}\!=\!j)/n^{(t,k)}, j=1,\cdots, C$.

In the original FedAVG \cite{mcmahan2017communication}, local models are averaged by the number of their devoted data samples to get the global model. In this case, $ \omega_t^f = \frac{n^{(t,k)}}{\sum_{k\in \mathcal{S}^t}n^{(t,k)}} \omega_t^k $. In this work, we borrow the idea of virtual client in FedVC \cite{hsu2020federated} as an auxiliary, in which clients with large datasets are separated while clients with small datasets will duplicate their samples, finally reaching a situation where all virtual clients have a dataset with size $N_{VC}$. Then the optimization steps taken by each virtual client are the same and all clients have the same weights in the aggregation process. FedVC has been validated to achieve better performance when there exists size imbalance among clients. Thus the aggregation method in our system is expressed as (\ref{equ-1}):
\begin{equation}
\label{equ-1}
\omega_t^f = \frac{1}{|S^t|}\sum_{k\in S^t} \omega_t^k.
\end{equation} 

All `client' refers to the `virtual client' in the remaining paper.

\subsection{Mathematical Demonstration}
{\color{black}We consider the classification problems and we use cross-entropy as the loss function.} Zhao et. al~\cite{zhao2018federated} proved that the non-IID settings in federated learning will cause weight divergence, which is the divergence between weights from the decentralized learning and weights from the centralized learning. The authors point out that the EMD between the data distribution on each client and the participated data distribution (population distribution) in each round is the root cause of the weight divergence. 

We extend the proposition in \cite{zhao2018federated} into a more general case, where the global data distribution is imbalanced and we define the new weight divergence as the difference between the weights obtained by the FedAVG algorithm $\omega_t^f$ and the optimal weights $\omega_t^*$ on the test dataset. The distribution of the test dataset is uniform among categories. To simplify the expression, the index $t$ in the superscript is omitted in the mathematical demonstration, which means that we neglect the dataset discrepancy through different rounds. The index $t$ in the subscript is rewritten as $mT$, where $T$ represents the optimization step conducted by each client at each round.

Given $|\mathcal{S}|=K$ clients with each client $k$'s local dataset following distribution $p_l^k$, there are $KN_{VC}$ samples in total. 
%Suppose $\nabla_\omega \mathbb{E}_{\mathrm{x}|y=i}[-\log f_i(\mathrm{x},\omega)]$ is $\lambda_{\mathrm{x}|y=i}$-Lipschitz for each class $j\in [C]$. 
{\color{black}We introduce an intermediate variable $\omega_t^c$, which physically represents the weights trained over the data from the selected clients in a centralized manner, to assist the derivation. Suppose $\nabla_\omega \mathbb{E}_{\mathrm{x}|y=i}[-\log f_i(\mathrm{x},\omega)]$ is $\lambda_{\mathrm{x}|y=i}$-Lipschitz for each class $j\in [C]$. Then we have the weight divergence bounded in (\ref{equ-3}). The boundary is obtained by summing up the weight divergence introduced in each optimization step using induction.}

\begin{equation}
\label{equ-3}
    \begin{split}
        ||\omega_{mT}^f-\omega_{mT}^*||
        \leq& ||\omega_{mT}^f-\omega_{mT}^c + \omega_{mT}^c-\omega_{mT}^*|| \\
        \leq& \frac{1}{K}\sum_{i=1}^K [(1+\eta \lambda)^T||\omega_{(m-1)T}^k - \omega_{(m-1)T}^c||\\
        &+\underbrace{\eta||p^k_l\!-\!p_o||_1}_{\textcircled{1}}(\sum_{j=2}^T \mathbf{g}(\omega_{mT-j}^c) (1\!+\!\eta \lambda)^{j-1})]\\
        &+(1+\eta \lambda)^T||\omega_{(m-1)T}^c - \omega_{(m-1)T}^*||\\
        &+ \underbrace{\eta||p_o\!-\!p_u||_1}_{\textcircled{2}}(\sum_{j=1}^T \mathbf{g}(\omega_{mT-j}^*) (1\!+\!\eta \lambda)^{j-1}),
        %\sum_{k=1}^K \sum_{j=1}^C||p_l^k(y=j)-p_o(y=j)|| \\&+\sum_{j=1}^C||p_o(y=j)-p_u(y=j)||
    \end{split}
\end{equation}
{\color{black}where $p_o$ is the population distribution, defined as the distribution of data participated at each round and $p_o(y\!=\!j)=\sum_{k\in \mathcal{S}} p^k_l(y\!=\!j)/|S|$; $p_u$ is the uniform distribution and $p_u(y\!=\!j) = 1/C$; $\eta$ is the learning rate and $\mathbf{g}(\cdot)$ is a function of $\omega$. The integrated derivation is online\footnote{\url{https://github.com/ICPP2021/Dubhe}}.} As shown in (\ref{equ-3}), we can observe that the weight divergence is from two aspects: (a) the weight divergence inherited from the last round; (b) the newly generated divergence from the data used in this round, which has two terms with term \textcircled{1} proportional to the EMD between the local data distribution and the population distribution and term \textcircled{2} proportional to the EMD between the population distribution and the uniform distribution.

The EMD between the data distribution on client $k$ and the population distribution is formally expressed as $EMD^k = ||p^k_l-p_o||_1$. $EMD^k$ is a characteristic of the local dataset on client $k$ and has no dependency on the client selection method. However, client selection methods are able to have significant influences on $||p_o - p_u||_1$. To ameliorate the impact of weight divergence and prevent skewed data to ``pull'' weights to other directions, the population distribution in each round should be uniform. 
In this way, the current problem can be expressed to be (\ref{equ-4}):
\begin{equation}
\label{equ-4}
    \mathrm{min} ||p_o - p_u||_1.
\end{equation}

In Dubhe, the $||p_o - p_u||_1$ can be reduced by up to $64.4\%$ in the worst case compared with the random client selection method.
%\added[]{Any final conclusion in one sentence/paragraph based on the above Equations??? The conclusion should support the design of Dubhe.}
\begin{figure*}
\centering
\includegraphics[width=0.9\textwidth]{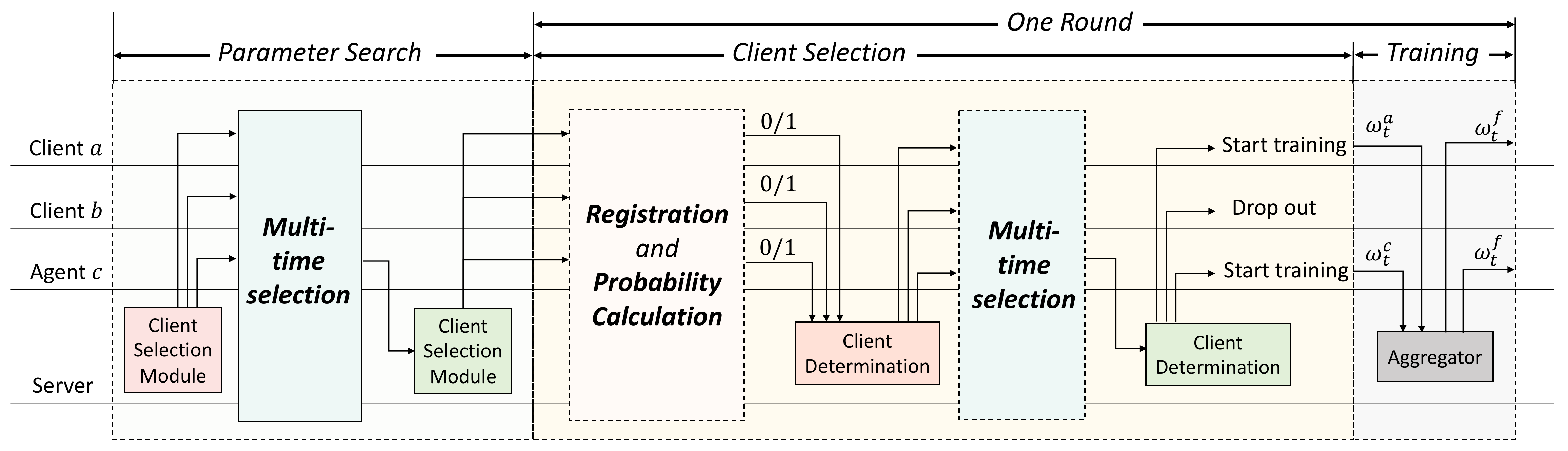}
\caption{The framework of Dubhe. Suppose there are three clients $a,b$ and $c$ while $c$ also plays the role of the agent. Client $a$ and $b$ finally participate in training in this round. Note that the client selection process and the training process in different rounds can actually be performed in parallel. Modules in dashed boxes are optional and modules in red refer to unsettled stages while green refers to settled.}
\label{fig:picture001}
\end{figure*}

\section{The design of Dubhe}
%\added[]{Why talk about FedAVG in the first sentence of Dubhe design?}

Dubhe deliberately selects clients according to the data distribution information to close the gap between $p_o$ and $p_u$.
%In FedAVG, the ratio of clients selected is fixed and the clients are selected randomly. Under this condition, $\mathbb{E}(p_o) = \mathbb{E}(\sum_{k \in \mathcal{S}} p_l^k)$. Astraea \cite{duan2020self} searches among clients one by one to achieve $\mathbb{E}(p_o) = p_u$, requiring the server to obtain the data distribution of all clients, which is strongly against the demand of privacy-preserving in FL. 
Figure~\ref{fig:picture001} shows a complete FL round with {\bf Dubhe} that consists of client selection and training.
In Dubhe, the data distribution information is encoded in a homomorphically encrypted structure, called registry. Each client participates in training with a probability computed by itself according to the registry. 
Dubhe requires a parameter search procedure to find the optimal parameters. The parameter search is performed whenever current parameters are not suitable for the FL system and a structural update for the system is required. 

There are several technical challenges in Dubhe. First, the encryption scheme in Dubhe is supposed to be designed delicately to avoid much encryption overhead. Second, clients in Dubhe should be autonomous and under independent control. Third, Dubhe should be robust and tolerant to the variations in the FL system (e.g., clients' data, the system capacity, the participation rate). 

There are three main components in the design of Dubhe to resolve the above challenges: 1) Registration (Figure~\ref{fig:registration}, Section~\ref{registration}), which encapsulates and memorizes the data distribution information of each client by its dominating data classes; 2) Probability calculation (Section~\ref{calculation}), which allows clients to proactively participate in training and compute their participation probability based on the registry, thereby balancing the population distribution in each round; 3) Multi-time selection (Figure~\ref{fig:picture002}, Section~\ref{search}), which can help to approach the uniform data distribution through repeated client selections and enables the parameter search function in a time-varying FL system. 

%\added[]{Which techniques are used to resolve the challenges?}

%\added[]{A processing step with Dubhe can be added.}

%\added[]{We should write the paper in a problem driven way, instead of describing the techniques directly. What is the naive method. What is the caused problem. Why our solution is able to solve the problem.}

\subsection{Registration}
\label{registration}
\begin{figure}
\centering
\includegraphics[width=0.47\textwidth]{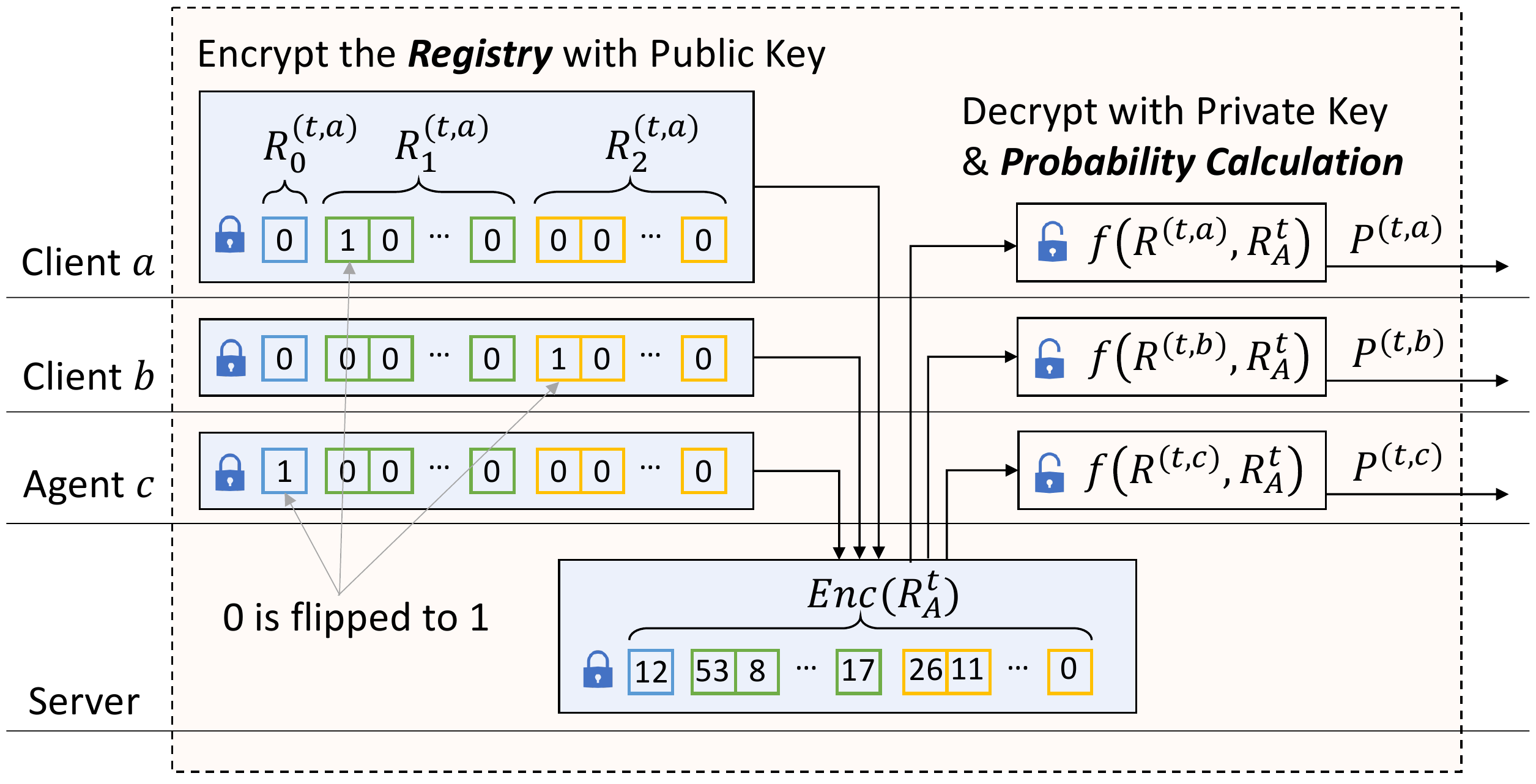}
\caption{Registration and probability calculation. In this example, we illustrate the case $G=\{1,2, 10\}$. It is also the method applied in the MNIST and CIFAR10 experiments.}
\label{fig:registration}
\vspace{-4mm}
\end{figure}

Figure~\ref{fig:registration} shows the registration process in Dubhe. As shown in the figure, each client joins the FL system by filling in a {\bf registry}. The registry is an all-zero vector before each client fills it by flipping only one specific zero to one. In this way, the registry encodes the client's data distribution information in a one-hot manner. The registry is then homomorphically encrypted and shared between clients and the server.

%\added[]{We may refer the part in the figure? }

The registration process is performed periodically in order to follow up on the states of clients. {\color{black}For the registration within round $t$, a HE key-pair $(\mathtt{pk}^t, \mathtt{sk}^t)$ is first generated and dispatched to all clients by a randomly selected agent to ensure security. Each client $k$ fills the registry $R^{(t,k)}$ according to its own data distribution, encrypts the registry with the public key $\mathtt{pk}^t$, and then transmits the encrypted registry $\mathtt{pk}^t(R^{(t,k)})$ to the server. After receiving all clients' encrypted registries, the server adds all the registries together and synchronizes the result to all clients. With the property of HE, the overall registry $R_\mathrm{A}^t$ can be decrypted by all clients who own the secret key $\mathtt{sk}^t$, as shown in (\ref{HE}):

\begin{equation}
\label{HE}
R_\mathrm{A}^t= \mathtt{sk}^t\left(\mathtt{pk}^t\left(R_\mathrm{A}^t\right)\right) =\mathtt{sk}^t\left(\sum_{k\in \mathcal{S}^t}\mathtt{pk}^t\left(R^{(t,k)}\right)\right).
\end{equation}

Each client $k$ can further proactively compute a probability $P^{(t,k)} = f(R^{(t,k)}, R_\mathrm{A}^t)$ according to the overall registry and its own registry.}

%There are several principles in the design of the registry. The registry should be able to be manipulated on the homomorphically encrypted space, in which case only the one-hot encoding method is both additive and feasible to represent the data distribution of clients.
One challenge problem here is that the size of the registry is significant. First, the encryption over the plaintext will expand the size of the registry, which is needed to be transmitted between clients and the server. Second, the sparsity of the overall registry can increase drastically with the registry size, further degrade the data balancing effect. 
In this case, the codebook of the registry should be designed carefully since we want the registry to conclude categories of data distribution as efficiently as possible.
%The registry can only reveals the data distribution information given a prescient dictionary and a registry's stored information size is in direct proportion to its length. Thus, it is significant to design a dictionary to fully represent each client's data distribution, since data balancing totally relies on the information given by registries.

%It is convenient to design a codebook with the priori knowledge of clients' profiles. For example, if we have already known that each client only has one class of data, then the registry can be set with a length of $C$, with each position $i\in [C]$ in the registry represents the client with data of class $i$. However, the design of the registry should also consider any priori access to global clients' profile unnecessary. 

%All these requirements impulse us to design a highly efficient registry. 
To resolve the above problem, the registry in Dubhe encodes each client's data distribution by its \textit{\textbf{dominating classes}} in classification problems. {\color{black}We use dominating classes to reduce the length of the one-hot encoded vector regarding the local data distribution. In this way, the overhead of handling the information with HE is reduced.}

The registry is generated according to a pre-determined reference set $G\subset [C]$ which contains possible numbers of dominating classes. We provide the general form of the registry as (\ref{equ-5}), which is a vector concatenated by several sub-vectors:
\begin{equation}
\label{equ-5}
R^{(t,k)}=\left[[R_i^{(t,k)}]\right], i\in G
%R^{(t,k)}=[R_1^{(t,k)}\cdots R_i^{(t,k)} \cdots R_e^{(t,k)}], i,e\in G,G\subset [C]
\end{equation}

Each client is categorized by its dominating classes as well as the number of dominating classes $i$, distinguished by the thresholds $\sigma_i$. Each sub-vector $R_i^{(t,k)}\in \{0,1\}^{l_i}$, while $l_i=\mathrm{dim}(R_i^{(t,k)})=C_c^i$ which is a combination number. The overall length of the registry $l=\sum_{i\in G} l_i$. For example, in the classification problem of MNIST dataset, we set $G = \{2, 10\}$ which presumes that there are 2 dominating classes or 10 dominating classes (equivalent to none of the classes dominates) in each client's dataset. In client $k$'s dataset, the data proportions of two dominating classes `$0$' and `$1$' both exceed $\sigma_2$, then the one's position in the registry should be in $R_2^{(t,k)}$, exactly at the place that represents class (0, 1). We denote $u^{(t,k)}$ as the \textit{\textbf{category}} of the client, while in this case $u^{(t,k)} =$ (0, 1). 
%The set of $u^{(t,k)}$ is $U^t$. 
The specific registration algorithm is shown in Algorithm \ref{algorithm}.
\begin{algorithm}
    \caption{Registration for client $k$ at round $t$}
    \label{algorithm}
    \KwData{$p^{(t,k)}, G, \sigma_i$ for $i\in G$}
    \KwResult{$R^{(t,k)}, u^{(t,k)}$}
    $m\leftarrow \text{zero vector of length |G|}$\;
    \For{$i$ in $G$}{
        \For{$j\leftarrow 1$ \KwTo $i$}{
            $u^{(t,k)}_j \leftarrow \mathrm{argmax}_{r\in [C]}(p^{(t,k)}(r))$\;
            $m_j \leftarrow p^{(t,k)}(u_j), p^{(t,k)}(u_j^{(t,k)})\leftarrow-1$\;
            \If{$m_i\geq \sigma_i$}{
                $R_i^{(t,k)}(u^{(t,k)}) \leftarrow 1$\;
                \Return{$R^{(t,k)}, u^{(t,k)}$}
            }
        }
    }
\end{algorithm}

Algorithm \ref{algorithm} presumes that the client has only one dominating class (extremely imbalanced) and check if the proportion of this dominating class exceeds $\sigma_1$, if failed then assumes there are two dominating classes and finally no dominating class (balanced).

\subsection{Probability Calculation}
\label{calculation}
The overall registry is $R_\mathrm{A}^t = \sum_{k=1}^N R^{(t,k)}$ and clients calculate their participation probability in (\ref{equ-6}).
\begin{equation}
\label{equ-6}
    \begin{split}
        P^{(t,k)}= f(R^{(t,k)}, R_\mathrm{A}^t) = \mathrm{min} \left(1, \  \frac{K}{R_\mathrm{A}^t(u^{(t,k)}) ||R_\mathrm{A}^t||_0}\right)
    \end{split}
\end{equation}

In (\ref{equ-6}), $R_\mathrm{A}^t(u^{(t,k)}) = R^{(t,k)} (R_\mathrm{A}^t)^\mathbf{T}$. Note that since the minimum possible value of $R_\mathrm{A}^t(u^{(t,k)})$ is $1$, we can eliminate the possibility that $P^{(t,k)}$ reach $1$ by restricting $K<||R_\mathrm{A}^t||_0$. With such participation probability, the expected participated number of clients in each round is fixed as expressed in (\ref{equ-7}).
\begin{equation}
\label{equ-7}
    \begin{split}
        \mathbb{E}(|\mathcal{S}^t|)= \sum_{k=1}^N P^{(t,k)} 
        = \sum_{u\in U^t}\frac{\mathbb{I}(R_\mathrm{A}^t(u)\neq 0) R_\mathrm{A}^t(u)\cdot K}{R_\mathrm{A}^t(u) ||R_\mathrm{A}^t||_0} = K
    \end{split}
\end{equation}

Besides, the expected counts of clients in each category are the same. As a qualitative explanation, it is obvious that if all categories in the combination set are chosen with the same probability, then the frequency of occurrence of each class as a dominating class is the same, stimulating the proportion of each class to be even. 
%The experimental results of this part can be found in Section \ref{}. 
The count of clients in category $u$ is $\kappa (u)$, whose expectation is calculated in (\ref{equ-8}).
\begin{equation}
\label{equ-8}
    \mathbb{E}[\kappa(u)] = \mathbb{E}[\sum_{k=1}^N \mathbb{I}(u^{(t,k)}=u)] = \frac{R_\mathrm{A}^t(u)\cdot K}{R_\mathrm{A}^t(u) ||R_\mathrm{A}^t||_0} = \frac{K}{||R_\mathrm{A}^t||_0}
\end{equation}

In Dubhe, we expect that the number of participated clients in each round is fixed to $K$. Thus, if $|\mathcal{S}^t|\leq K$, we uniformly select $K-|\mathcal{S}^t|$ clients to replenish $\mathcal{S}^t$, otherwise we uniformly remove $|\mathcal{S}^t|-K$ clients from $\mathcal{S}^t$.

\subsection{Multi-time selection}
\label{search}
\begin{figure}
\centering
\includegraphics[width=0.47\textwidth]{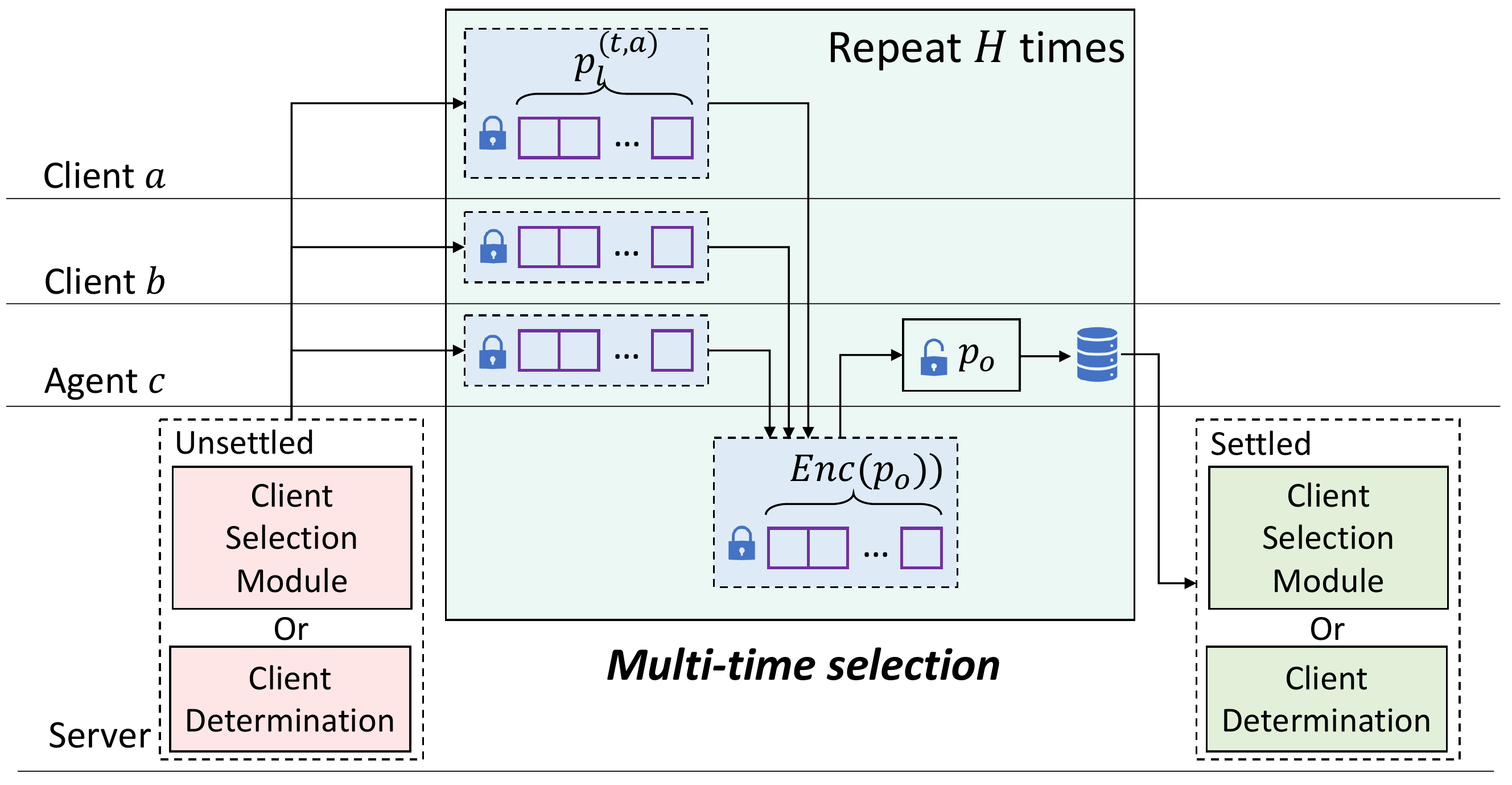}
\caption{The multi-time selection process. Note that the inputs and outputs are different for the parameter search process and the client determination process.
%The client selection process can be repeated for several times when the client selection module is fixed, in order to reduce variance and improve authority in the parameter search process.
}
\label{fig:picture002}
\vspace{-4mm}
\end{figure}

With the property of HE, we can check the distance between $p_o$ and $p_u$ by frequently exchanging information between clients and the server. Figure~\ref{fig:picture002} illustrates the essence of this multi-time client selection. This process improves data unbiasedness by conducting tentative client selections repeatedly. The process helps to further reduce $||p_o - p_u||_1$ in client determination and improve the credibility of parameter search.

In a multi-time selection with $H$ tentative tries, the clients selected in each tentative try $h$ send the encrypted distribution $\mathtt{pk}(p_l^k)$ to the server. Then the server sends back the aggregated distribution $\sum_{k\in \mathcal{S}} \mathtt{pk}(p_l^k) = \mathtt{pk}(\sum_{k\in \mathcal{S}} p_l^k) = \mathtt{pk}(p_o)$ to an agent, which is a client randomly selected by the server. The agent records the $p_o$ in each try $h$ as $p_{o,h}$. The $H$ tries in the multi-time selection can be conducted in parallel, without adding much encryption and communication overhead.

\subsubsection{Client Determination}
%We have explained that the proposed client-selection algorithm can reach $\mathbb{E}(p_o)=p_u$ though with skewness to some extent.
The participated clients are determined by the server and the most natural client determination solution is to do the one-off determination each round. In Dubhe, a $H$-time selection is used to determine the best clients set within $H$ tentative selections in each round, thereby balancing data in an efficient way. The specific tentative selection algorithm with a $H$-time try is shown as follows: at the $h_{th}$ try, the set of clients selected by the server is denoted as $\mathcal{S}^t_h$ and is recorded by the server. After $H$ tries, the agent finds the optimal try $h^* = \mathrm{argmin}_h (||p_{o,h}-p_u||_1)$ and the ultimate determined clients are in $\mathcal{S}_{h^*}$.

%However, frequent information switch between clients and the server is communication heavy and there is a trade-off between accuracy and communication overhead.

\subsubsection{Parameter Search}

{\color{black}The parameter search in Dubhe is to find proper parameters (thresholds) for the registration. Whenever the systematic structures of the FL system (e.g., the global data pattern, the total client number, the participation rate) are changed drastically, making the current parameters inapplicable, the parameter search is performed to update the parameters.

The threshold $\sigma_C$ can be determined directly because $i\!=\!C$ is a special element that must be included in the reference set $G$. The corresponding vector $R_C^{(t,k)}$ is a special vector with size 1. Any client who fills in this vector indicates that there are no dominating classes in its dataset, which can be regarded as a dataset that is rather balancing. Thus, according to Algorithm \ref{algorithm}, we have $\sigma_C=0$.

In each step in the parameter search, the unsettled client selection module picks a set of parameters $\sigma_i, i\in G$ from the parameter space and sends them to all clients. The registry form and its corresponding codebook are also dispatched. Clients then register and calculate their participation probability according to the given parameters. For each fixed set of parameters, $\mathbb{E}_h(p_{o,h})$, which is the expectation of $p_o$, is obtained within $H$ tentative tries. The server then traverses the parameters in the parameter space. Ultimately the optimal set of parameters that minimize $||\mathbb{E}_h(p_{o,h})-p_u||_1$ is provided by the agent without any information leakage to the server. The optimal set of parameters is applied and the client selection module is settled.}

\section{Evaluation of Dubhe}

\begin{figure*}
\begin{minipage}[t]{0.71\textwidth}
\centering
\includegraphics[width=1\textwidth]{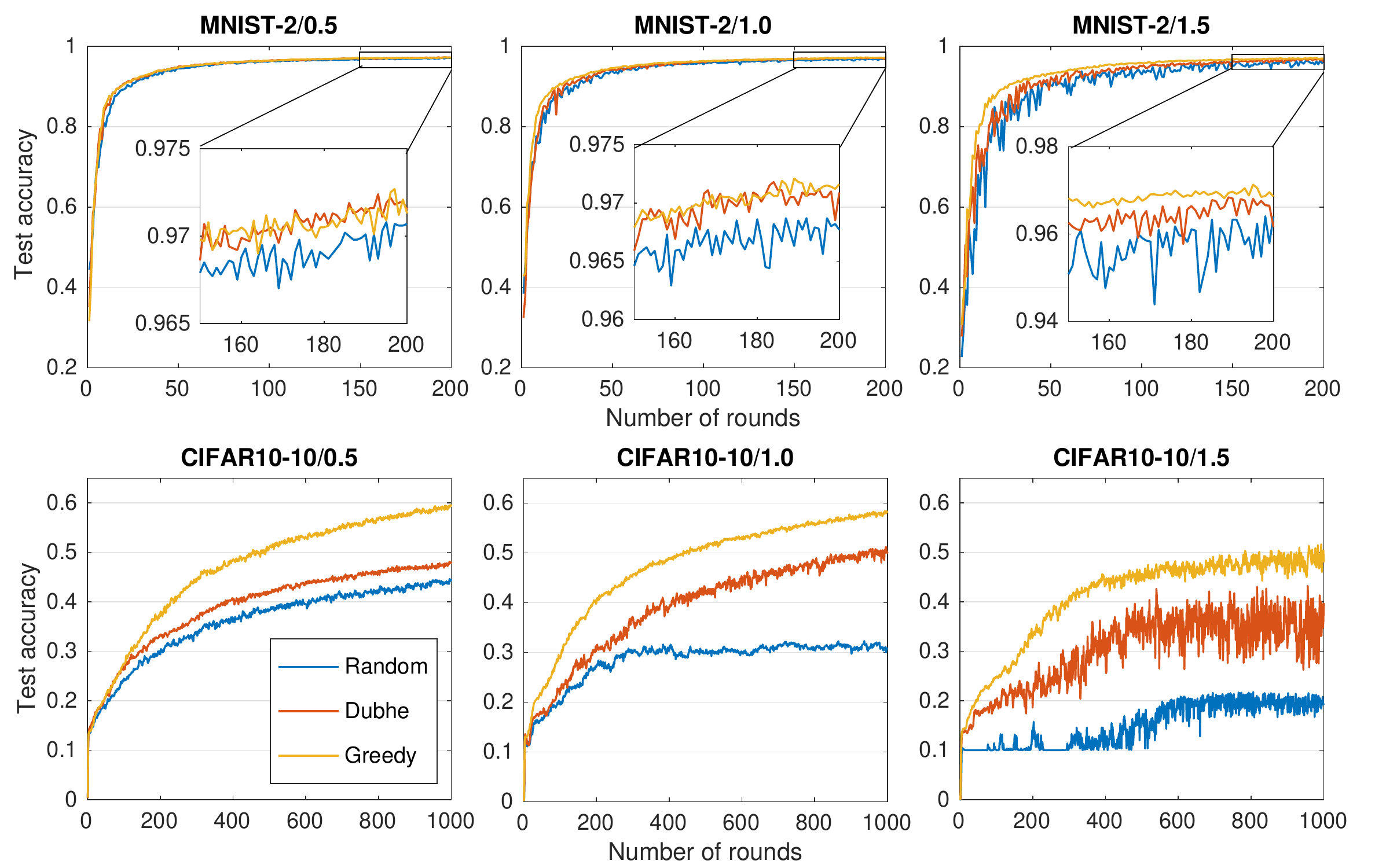}
\caption{Test accuracy curves on MNIST-2/$EMD^{avg}$ and CIFAR10-10/$EMD^{avg}$. \\$EMD^{avg}$ = $\{0.5, 1.0, 1.5\}$.}
\label{fig:cross-left}
\end{minipage}
\begin{minipage}[t]{0.243\textwidth}
\centering
\includegraphics[width=1\textwidth]{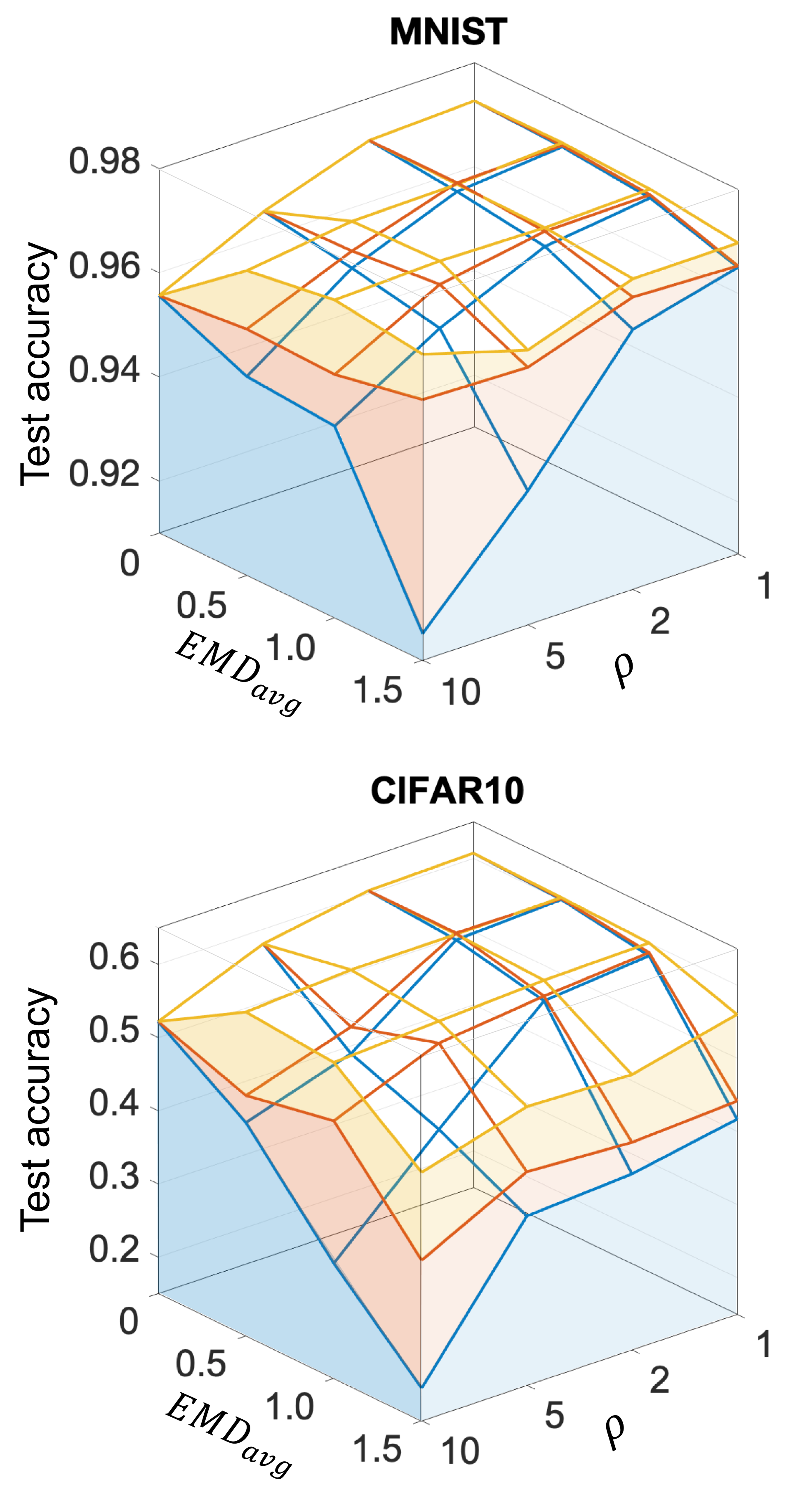}
\caption{Average accuracy over the last 50 rounds.}
\label{fig:cross-right}
\end{minipage}
\vspace{-3mm}
\end{figure*}

%\added[]{In this section, we xxx.}
In this section, we first introduce the experimental setup, then evaluate the effect of accuracy in various datasets. We analyze the data unbiasedness achieved by Dubhe and its performance including the encryption and communication overhead.

\subsection{Experimental setup}
We conduct our experiments based on \texttt{FedML}~\cite{chaoyanghe2020fedml}, a research library for FL with Pytorch 1.8.0 and CUDA 11.2. We select different data partitions to form dataloaders in each round and implement the training process of participated clients as parallel processes. The homomorphic encryption is based on the open-sourced \texttt{Python Paillier} \cite{PythonPaillier}.

We apply three algorithms throughout our experiments, which are the random selection (as our baseline), the greedy selection (as the optimal bound) and Dubhe. In the greedy selection, the server first randomly selects a client, then continuously select clients to make the KL divergence from the data distribution of selected clients to the uniform data distribution to be minimum. Readers can refer to \cite{duan2020self} for more details. The greedy selection method requires an overall knowledge of clients' data distribution, which is not applicable in secure FL. {\color{black}Besides, we also observe a $0.13\times$ additional client selection time compared with the total elapsed time when $N=1000$ and $1.69\times$ when $N=8962$, due to the high time complexity of greedy selection.}
%\added[]{Describe the drawback of greedy selection. It backs up the decision of not using greedy selection.}

\subsubsection{Dataset Generation}
We use two series of datasets generated from MNIST and CIFAR10. We simulate the imbalanced property of data by sampling datasets with half-normal distributions~\cite{cui2019class}. In the two series of synthetic datasets, the class imbalance ratio $\rho$ is used to control the skewness of global data and is defined as the sample size of the most frequent class divided by that of the least frequent class within overall data.

{\color{black}We also adopt the Federated Extended MNIST (FEMNIST) \cite{caldas2018leaf} dataset which consists of 10 classes of handwritten digits (MNIST) and 52 classes of handwritten letters. In our experiment, FEMNIST, with original 3400 data partitions is further split to 8962 clients with an even number of samples. We use the letters dataset and classify over the 52 classes of letters.}

The average Earth Mover's Distance $EMD^{avg}=\sum_{k=1}^N EMD^k$ is used to evaluate the discrepancy of data distribution among all clients~\cite{zhao2018federated}. %\added[]{For $EMD^{avg}$, the larger the more discrepancy of the data distribution?} 
We explain this discrepancy by introducing two extreme cases. In the first case, all clients have the same data distribution with the global data distribution ($EMD^{avg}=0$). In the second case, each client only has one category of data ($EMD^{avg}$ is maximized). We generate distributions with properties between those two extreme cases. The datasets used in our experiments are shown in Table. \ref{table00}. We name datasets by ``{Dataset\_Name}-$\rho/EMD^{avg}$'', e.g., ``CIFAR10-2/1.0''.
%The dataset generation algorithm is shown in Appendix X.
%Virtual clients are established as different Pytorch dataloaders and sample data from the overall dataset round by round following the preset distributions in Table X.

\begin{table}[]
\caption{The datasets used in our experiments.}
\centering
\begin{tabular}{c|c|c|c}
\hline
Dataset & imbalance ratio $\rho$    & $EMD^{avg}$  & $N$ \\ \hline
MNIST & \multirow{2}{*}{$10,5,2,1$} & \multirow{2}{*}{$0.0,0.5,1.0,1.5$} & \multirow{2}{*}{1000}  \\ \cline{1-1}
CIFAR10 &        &       &      \\ \hline
FEMNIST & 13.64 & 0.554 & 8962 \\ \hline
\end{tabular}
\label{table00}
\vspace{-2mm}
\end{table}

\subsubsection{Configurations}
We separate our experiments by the class number of datasets. As the first group in our experiments (MNIST and CIFAR10), the class number $C=10$ and the training parameters are $B = 8, N_{VC}=128, E=1, K=20$. As the second group of experiments (FEMNIST), the class number $C=52$ and the training parameters are $B = 8, N_{VC}=32, E=5, K=20$. For local training, each client updates the weights via Adam optimizer with learning rate $lr=1\mathrm{e}-4$ and no weight decay. As for MNIST and FEMNIST, the model used are the CNNs proposed in \cite{reddi2020adaptive}, while in the experiment of CIFAR10, we use Resnet18 for training.

%The total number of clients $N$ is 1000 and number of clients selected each round is $20$ in the basic experiment. In each round, each client pick $128$ data samples from the global dataset with $p_l^k$. 
Considering the scale of the classification problem in experiments, we set the reference set $G^1=\{1,2, 10\}$ in group 1 and $G^2=\{1, 52\}$ for group 2, then the length of the registry in group 1 is $l^1=C_{10}^1+C_{10}^2 + C_{10}^{10}=56$ and $l^2=C_{52}^1+C_{52}^{52}=53$ for group 2, which are both compatible to the client capacity (1000 and 8962) of the FL systems.

\subsection{Training Accuracy with Dubhe}

With different $EMD^{avg}$, Figure~\ref{fig:cross-left} shows the accuracy curves of MNIST with $\rho = 2$ and CIFAR10 with $\rho = 10$. For MNIST, the accuracy of Dubhe is very similar to that of the greedy selection method. There is also a remarkable improvement from the random selection method for CIFAR10. It is observed that the training curve contains more fluctuations with the increment of $EMD^{avg}$, because of the fact that when $EMD^{avg}$ is large, each client owns significantly different data and pulls the model to reveal its own data. There is also an observation that when $\rho$ is large, e.g., 10, the random selection method is more prone to lead the model to a local optimum. Notably, in CIFAR10-10/1.5, the accuracy is merely $0.1$ at the beginning which indicates that the model does not learn any information.

We examine the three client selection methods on datasets with different $\rho$ and $EMD^{avg}$ and the average test accuracy over the last 50 rounds are shown in Figure~\ref{fig:cross-right}. With the decreasing of the imbalance ratio and the increase of $EMD^{avg}$, the accuracy of classification models is decreasing using the random selection. Dubhe and the greedy selection have the same properties when $EMD^{avg}=0$ or $\rho=1$ because there is no room for algorithms to balance data in these situations. However, through filling the gap between the population distribution and the uniform distribution, Dubhe and the greedy selection method efficiently prevent the model from degrading when statistical heterogeneity is severe, typically when $EMD^{avg}=1.5$ and $\rho=10$.
%\added[]{add some explanation on what happens when $EMD_avg$ increses, when $\rho$ increases etc.}

\begin{figure}[t]
\centering
\includegraphics[width=0.48\textwidth]{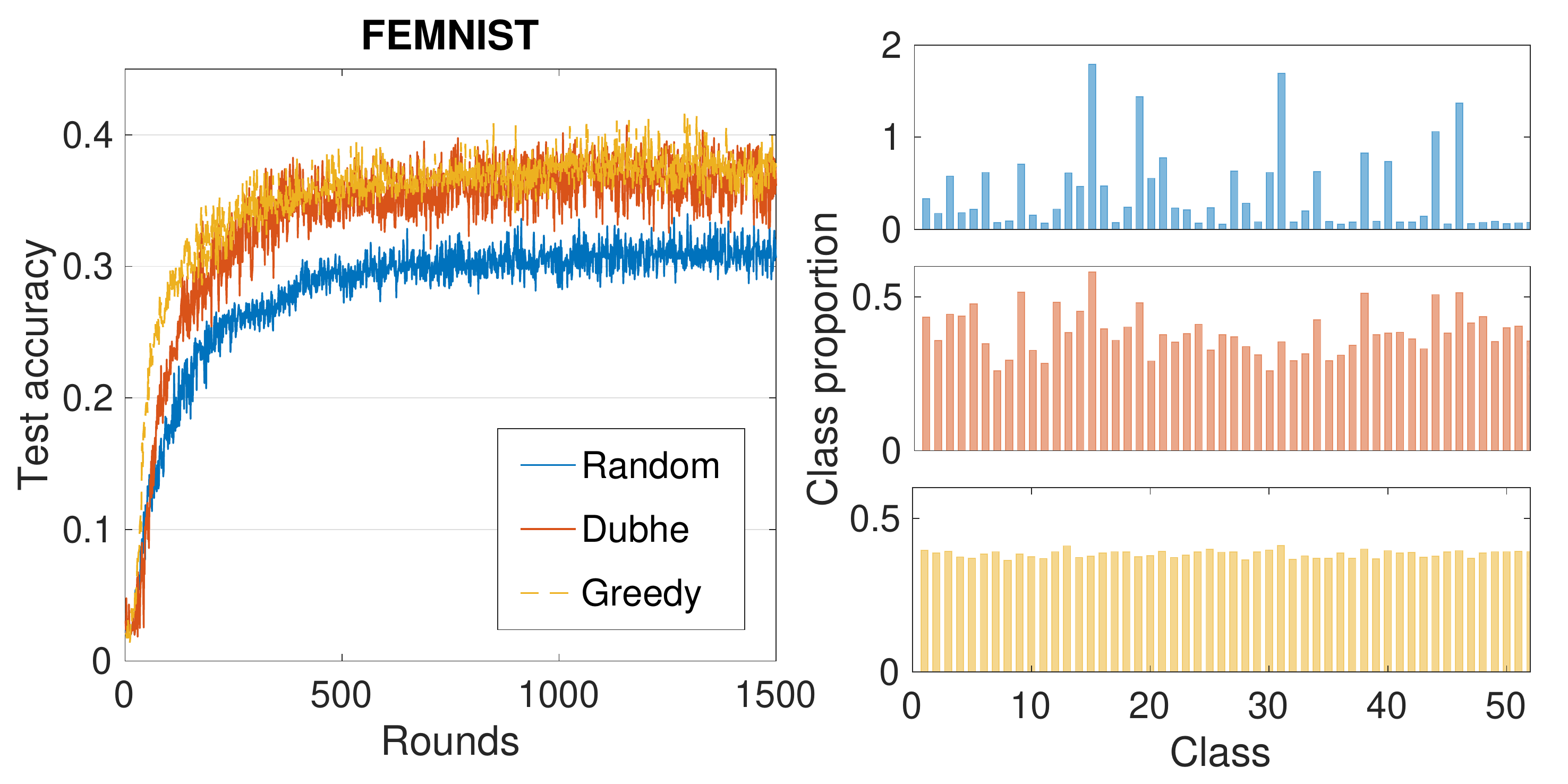}
\vspace{-5mm}
\caption{Results on FEMNIST. Left: the test accuracy curves of different methods. Right: the population class proportion in one random round.}
\label{femnist}
\vspace{-6mm}
\end{figure}

The results of FEMNIST are shown in Figure~\ref{femnist}, with the random selection achieve the test accuracy of $31.0 \%$, Dubhe of $36.4\%$ and the greedy selection of $37.4\%$. It is worth noticing that the convergence rate of the greedy selection is greater than the Dubhe's, then the random selection's. 

The test accuracy has a direct relationship with the population distribution. The population proportion of FEMNIST in one random round is shown on the right of Figure~\ref{femnist}. In random selection, the expectation of the population distribution over $\tau$ rounds is the same as the global data distribution since $\mathbb{E}_\tau (p_{o,\tau}) = \mathbb{E}_\tau(\sum_{k \in \mathcal{S^\tau}} p_l^k)$. The data balancing performance of Dubhe is approaching the performance of the greedy selection as shown in Figure~\ref{femnist}.

\subsection{Impact of Data Unbiasedness}
We have theoretically and experimentally verified that $||p_o-p_u||_1$ has a direct impact on the model accuracy. In this subsection, we discuss how system parameters and components in Dubhe influence $||p_o-p_u||_1$.

\subsubsection{System Parameter}
\begin{figure}[t]
\centering
\includegraphics[width=0.48\textwidth]{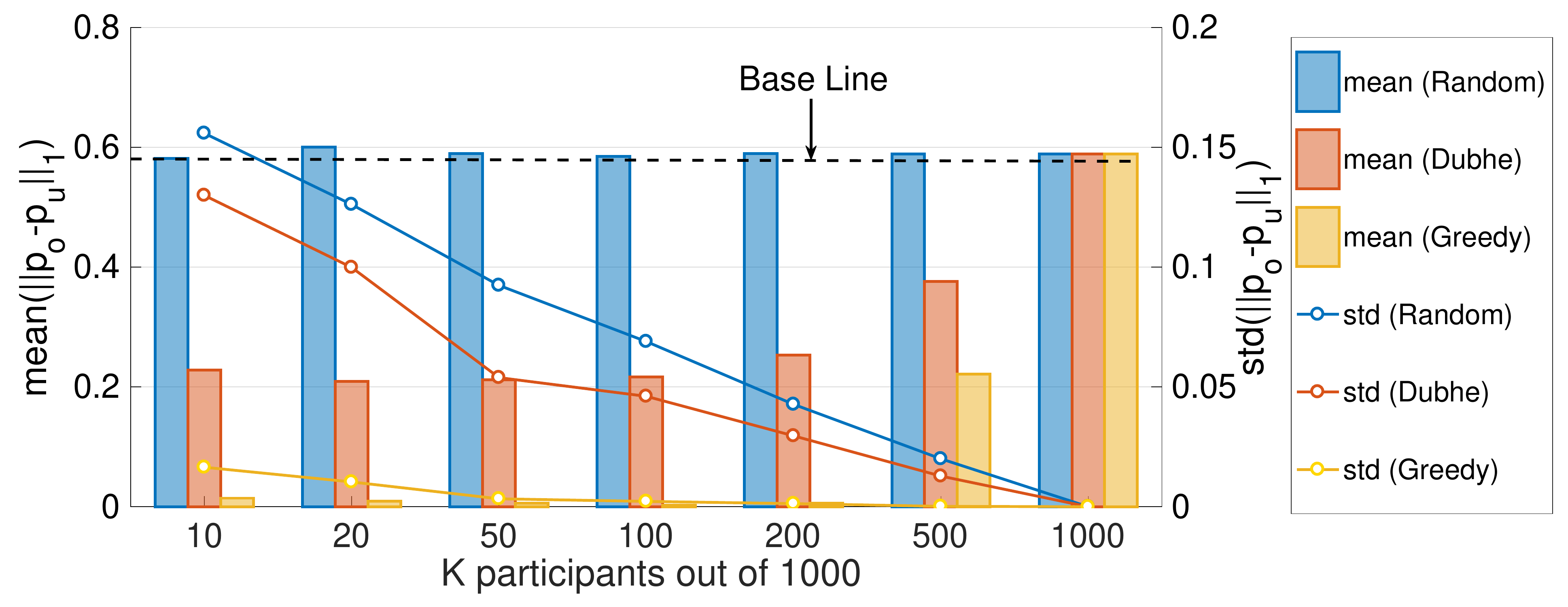}
\caption{The data unbiasedness with different methods. The dataset used here is MNIST/CIFAR10-10/1.5. The base line is $||p_g-p_u||_1$, where $p_g$ is the global data distribution.}
\label{fig:picture005}
\vspace{-4mm}
\end{figure}

Datasets from group one are used in this part for illustration and explanation. The imbalance ratio $\rho=10$ and $EMD^{avg}=1.5$, which are representative. The conclusion drawn with this kind of dataset is common over all possible datasets. There are total $1000$ clients in the FL system and the participation rate varies from $10/1000$ to $1000/1000$ in our experiment. As shown in Figure~\ref{fig:picture005}, we draw the average $||p_o-p_u||_1$ with histograms and the deviation of $||p_o-p_u||_1$ with lines over 100 times of selections.

Through random selection, the average population distribution is close to the global data distribution, but with a large standard deviation when the participation rate is small. The phenomenon reveals that when the global data distribution is skewed, the population distribution would always shift away from the uniform distribution, regardless of the participation rate. 

Through the greedy selection method, the participated data can reach unbiasedness perfectly when the participation rate is low. However, with the increase of the participation rate, the population distribution also begins to follow the global data distribution. %The three algorithms have no difference in client selection when the participation rate is approaching 1.

Through Dubhe, the discrepancy of the population distribution and the uniform distribution is highly suppressed when the participation rate is low, even when the global data is seriously skewed. The $||p_o-p_u||_1$ is reduced by $64.4\%$ with Dubhe compared with the random selection method when $\rho=10$ and $EMD^{avg}=1.5$. The decent robustness of Dubhe to the participation rate is revealed. There is also an observation that the standard deviation decreases with the increase of participation rate because larger participation reduces the bias in population distribution.

%With the client participation rate fixed, we have also tested the impact caused by the system scale.

\subsubsection{Multi-time Selection}
With the auxiliary of multi-time selection, clients can be lastly determined through a $H$-time selection instead of a one-off determination. The effect of $H$ on the data unbiasedness and model accuracy is shown in Table \ref{table01}. In the table, $EMD^*\!=\! ||p_{o,h^*}-p_u||_1$ and $\beta$ are the improvement of the accuracy compared with the single-time selection, while the accuracy improvement in the greedy selection (noted as opt) is considered as $100\%$. 
As observed in Table \ref{table01}, $EMD^*$ decreases with larger $H$, thereby improving the model accuracy. The improvement of MNIST can reach $69.5\%$ when $H=10$ and the improvement of CIFAR10 can reach $18.8\%$ when $H=20$. However, the accuracy improvement is not strictly proportional to $H$ due to the randomness in client selection. More times of tentative selection results in greater data unbiasedness and potentially higher accuracy.

\begin{table}[]
\centering
\caption{Results with multi-time selection. $M$ refers to MNIST and $C$ refers to CIFAR10.}
\label{table01}
\begin{tabular}{c|c|c|c|c|c||c}
\toprule[1pt]
H       & 1      & 2       & 5       & 10      & 20       & opt    \\ \midrule \midrule
$EMD^*$     & 0.2946 & 0.2588  & 0.2176  & 0.1971  & {\bf 0.1750}   & 0.0144      \\ \midrule
$Acc^M$  & 0.9662 & 0.9668  & 0.9665  & {\bf 0.9684}  & 0.9678   & 0.9694 \\ \midrule
$\beta^M$  & 0.0\% & 17.6\% & 10.5\% & {\bf 69.5\%} & 51.5\%  & 100\%  \\ \midrule
$Acc^C$ & 0.4300 & 0.4518  & 0.4486  & 0.4441  & {\bf 0.4577}   & 0.5295 \\ \midrule
$\beta^C$      & 0.00\% & 14.8\% & 12.6\% & 9.5\%  & {\bf 18.8\%}   & 100\%  \\ \bottomrule[1pt]
\end{tabular}
\vspace{-4mm}
\end{table}

\subsubsection{Registry Sparsity}
\begin{figure}[t]
\centering
\includegraphics[width=0.45\textwidth]{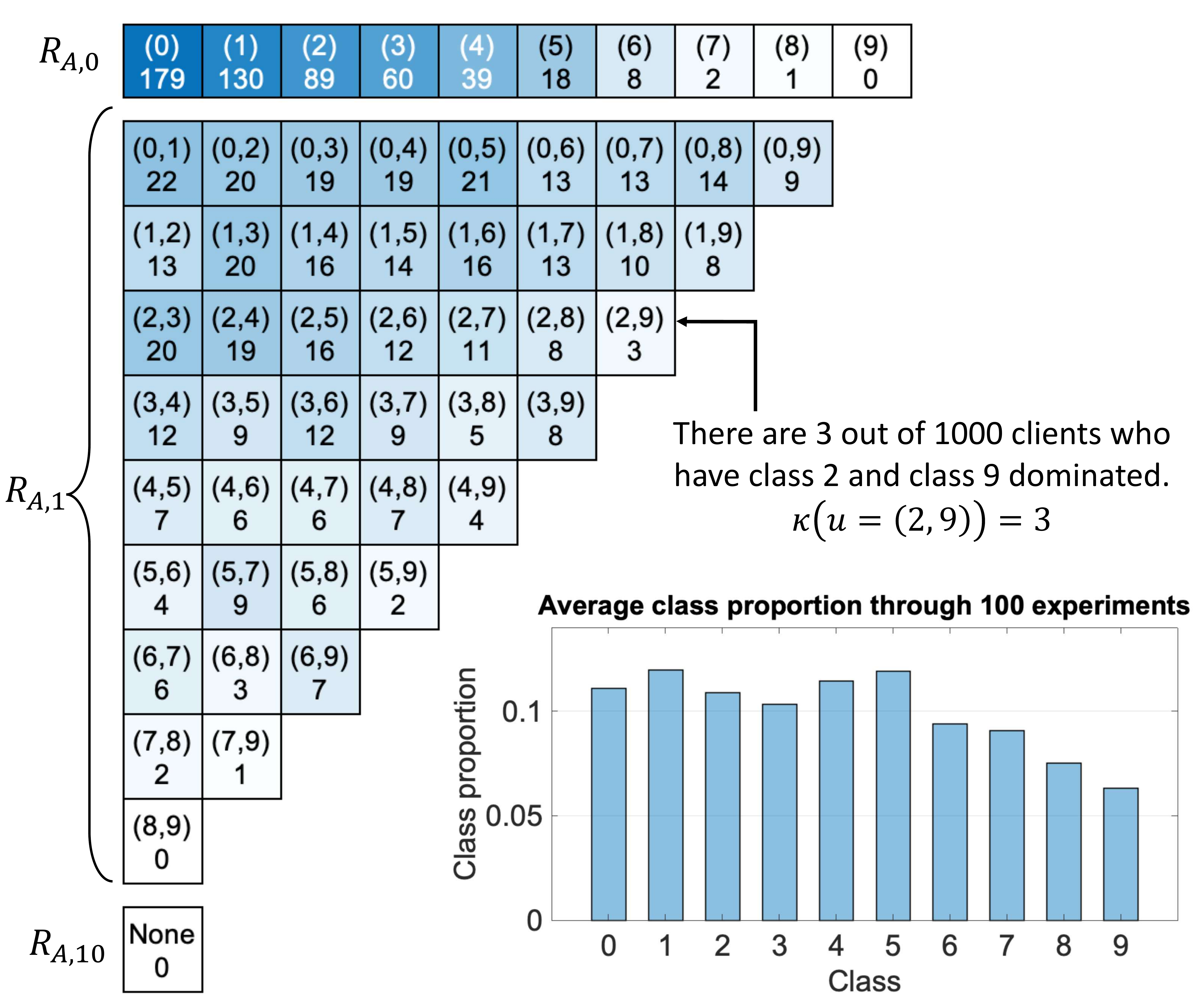}
\caption{An example of the overall registry and its corresponding participated class proportion. The category of clients and its count are recorded in each block.}%\added[]{How to understand each element, such as (0,1)22 in $R_1$?}
\label{fig:picture006}
\vspace{-4mm}
\end{figure}

The registry sparsity is another root cause of the skewness of data distribution. Figure~\ref{fig:picture006} illustrates the overall registry in group one datasets with $N=1000$, $EMD^{avg}=1.5$, $\rho=10$ and $G = \{1,2,10\}$. The optimal parameters found by the parameter search process are $\sigma_1 = 0.7, \sigma_2=0.1$. Under this setting of the FL system, we run the experiment for $100$ times and obtain the average population distribution as shown in Figure~\ref{fig:picture006}. Compared with the global class proportion where $\rho = 10$, the population proportion is rather balancing. However, the minority classes are still in inferior positions, where class 8 has a proportion of 0.0753 (which we expect to be 0.1) and the proportion of class 9 is 0.0632.

The sparsity of the overall registry is the primary cause of this deviation from expectation. As shown in Figure~\ref{fig:picture006}, except from the category in $R_{A,10}$ which refers to clients who have no dominating classes, the counts of client category $(9)$ and $(8,9)$ are also $0$. Thus, there are no clients who have class 9 dominated participated in training. The skewness of the global data distribution makes this phenomenon inevitable. Nevertheless, the registry sparsity problem can be alleviated with the increase of total number of clients. With a large population base, the number of minorities can also be considerable. The native local skewness on each client is another inescapable factor that causes $p_o$ to break away from $p_u$. For example, if samples of class 1 are less than samples of class 0 on all clients, then the registry will never consider class 1 a dominating class over class 0, which makes it impossible to balance class 1 with class 0 through client selection methods under this situation.

\subsection{Encryption and Communication Overhead}
The encryption and communication overhead of Dubhe is negligible compared with the basic model training and communication. The encryption overhead is proportional to the plaintext size in HE. As for Dubhe, we encrypt with the key of size 2048 in Paillier, which is adopted in \cite{FATE, zhang2020batchcrypt}. The overhead in Dubhe exists in the registration and the multi-time selection process.

{\bf Overhead in registration:} In our experiments, the lengths of the registries are $56$ and $53$. The corresponding plaintexts are $0.47-0.49$ KB large in Python3. After the encryption of \texttt{Python Paillier}, the sizes of the ciphertexts are expanded to $29.6-31.28$ KB. The encryption of a registry with size $56$ requires $6.9$ seconds and the decryption requires $1.9$ seconds on average. 
%When the data distribution is invariant, the encryption and the decryption process are conducted only once. 
Compared with the training process that takes hours, the encryption overhead in the registration is trivial. Since the registration and the training in different rounds can be conducted asynchronously, even frequent registrations will not hinder the training. 
%Compared with the encryption of weights, the overhead of the encryption of registries is rather small. 

{\bf Overhead in multi-time selection:} The multi-time selection requires the encryption on $p_l$. We consider the length of $p_l$ to be $C=52$ here. The size of such a plaintext is $0.68$ KB and the size of the ciphertext is $29.1$ KB. The encryption requires $6.8$ seconds and the decryption requires $1.7$ seconds on average. The encryption is operated in parallel on clients, and the multiple times of decryption are operated in parallel on the agent. 
%The encryption overhead in the parameter search process is not significant since it is only performed when a structural update of the FL system is required. 
However, we have to notice that if the multi-selection method is used for client determination, then approximately $(H-1)K$ additional clients have to be active and participate in the encryption and communication of their data distribution in each round.

Except for the overhead from the optional multi-time selection for client determination, the overhead of Dubhe only exists when updates are required (the registration is required when clients' data distribution varies and the parameter search is required when systematic structures are changed).

Encrypted registries and $p_l$ are negligible (KBs) compared with model weights (MBs or GBs~\cite{canziani2016analysis}) in terms of sizes. %Figure~\ref{fig:picture001} shows the extra communication times introduced by Dubhe. 
{\color{black}We then use the times of communication to measure the communication overhead.} In classic FL systems%~\cite{bonawitz2019towards}
, each participated client checks in with the server and the selection process requires $K$ times of communication in each round. Additional communication overhead in Dubhe comes from two parts. First, whenever there is a requirement of new registration, it requires $N$ times of communication transferring the ciphertext of the registry. Second, the multi-time selection process requires $HK$ times of communication approximately in each round before the client determination. 
%The communication cost for the client selection has no difference with the classic one when $H=1$.

%\added[]{We may add some discussion here if still have space.}

\section{Conclusions}
In this work, we mathematically demonstrate the impact of data skewness on the performance degradation in FL. Towards the unbiasedness of data in FL, we propose Dubhe, a proactive client selection system to balance skewed data. Dubhe is pluggable, adaptive and robust to various FL settings, with negligible encryption and communication overhead. By applying homomorphic encryption, Dubhe improves the training performance without bringing security threats. 
%The implementation of Dubhe is orthogonal with the training process and the cost of Dubhe is also negligible compared with the computation and communication in training. 
We have tested Dubhe under different skewed classification datasets and Dubhe achieves superior data unbiasedness compared with the random selection method and approaches the results of the optimal greedy selection method.

\begin{acks}
This work is partially sponsored by the National Natural Science Foundation of China (NSFC) (62022057, 61832006, 61632017, 61872240). Quan Chen and Minyi Guo are the corresponding authors.
\end{acks}

\bibliographystyle{ACM-Reference-Format}
\bibliography{acmart}

\end{document}